\documentclass{article}
\usepackage{graphicx}%
\usepackage{multirow}%
\usepackage{amsmath,amssymb,amsfonts}%
\usepackage{authblk}
\usepackage{hyperref}

\begin{document}

\title{Semileptonic decay and form factors of $\Omega_b^- \rightarrow \Omega_c^0\,e\,\bar{\nu_e}$ }

\author{Kinjal Patel \thanks{kinjal1999patel@gmail.com}}

\author{Kaushal Thakkar\thanks{Corresponding Author: kaushal2physics@gmail.com}}

\affil{Department of Physics, Government College, Daman-396210,
U. T. of Dadra $\&$ Nagar Haveli and Daman $\&$ Diu, Veer Narmad South Gujarat University, Surat, India }
\maketitle
\begin{abstract}We investigated the heavy-to-heavy semileptonic decay $\Omega_b^- \rightarrow \Omega_c^0 e \bar{\nu_e}$ within the framework of the Hypercentral Constituent Quark Model (HCQM). The ground-state masses of the involved baryons were evaluated by numerically solving the six-dimensional hyperradial Schr\"{o}dinger equation, incorporating both hyper-Coulomb and linear confinement potentials along with spin-dependent interactions. The Heavy Quark Effective Field Theory (HQET) form factors are computed up to the subleading order, incorporating $1/m_Q$ corrections that account for finite mass effects beyond the heavy-quark symmetry limit. These form factors were then employed to analyse the heavy-to-heavy semileptonic decay rate via helicity formalism. The decay width and branching ratio results were compared to those obtained using various theoretical approaches.
\end{abstract}


\section{Introduction}

The investigation of weak decays in heavy-flavoured hadrons has played a crucial role in the construction and verification of the Standard Model (SM).  In particular, the semileptonic decay of heavy-flavoured baryons provides valuable insights into their weak transition dynamics. Compared with mesonic decays, baryonic semileptonic decays offer additional information regarding the dynamics of heavy quarks bound in a three-body system. To date, more than 60 singly heavy baryon channels have been experimentally observed in charm and bottom sectors \cite{pdg2024}. These experimental discoveries have motivated a wide range of theoretical studies to understand their structural and decay properties. Theoretically, properties such as mass spectra, Regge trajectories, electromagnetic properties, and weak decay of heavy baryons have been extensively investigated using various theoretical approaches. The study of the semileptonic decay of heavy baryons provides an essential window into the nonperturbative regime of Quantum Chromodynamics (QCD). In particular, transitions governed by the $b \rightarrow c$ quark process allow for precise testing of the Heavy Quark Effective Theory (HQET) and offer an independent determination of the Cabibbo-Kobayashi-Maskawa (CKM) matrix element $|V_{cb}|$. \\

Among the bottom baryons, the $\Lambda_b$ and $\Xi_b$ have been the most extensively studied, both experimentally and theoretically, serving as benchmarks for understanding the weak transitions of heavy baryons. Currently, $\Lambda_b$ baryon is the only heavy baryon for which semileptonic decay has been experimentally observed. In contrast, investigations of the weak decay of heavy-flavoured baryons containing two strange quarks remain limited. The weak semileptonic channel $\Omega_b \rightarrow \Omega_c\ell\bar{\nu_{\ell}}$ has not yet been observed. The presence of two strange quarks in $\Omega_b$ gives rise to unique flavour and spin correlations, distinguishing the $\Omega_b$ from other bottom baryons and making its semileptonic transitions particularly important for testing theoretical models. Notably, the LHCb collaboration has already observed the production of the $\Omega_b$ baryon and its mass spectrum,  suggesting that investigating its weak decay is feasible in future experiments.

The $\Omega_b$ baryon was first observed at Tevatron, where the $D\emptyset$ and $CDF$ Collaborations reconstructed it through the weak decay channel $\Omega_b^- \rightarrow J/\psi \Omega^-$ with subsequent decays $J/\psi \rightarrow \mu^+ \mu^-$, $\Omega^- \rightarrow \Lambda K^- \rightarrow (p\pi^-)K^-$ \cite{Abazov2008,{Aaltonen2009}}. Subsequent measurements established the properties of the $\Omega_b^-$ baryon with improved precision. The Particle Data Group (PDG) \cite{pdg2024} quotes its mass as $M_{\Omega_b^-} = 6045.8 \pm 0.8 MeV$ and its lifetime as $\tau_{\Omega_b^-} = (1.64 \pm 0.16) ps$. These measurements provide important benchmarks for testing quark models and effective field theories, because the $\Omega_b$ is the only single-bottom baryon containing two strange quarks. Therefore, understanding its decay properties is essential not only for completing the spectroscopy of bottom baryons but also for extracting information about the underlying quark dynamics. The semileptonic transition of $\Omega_b$ baryons has been studied using various theoretical models such as the QCD sum rule \cite{Neishabouri2024,Khajouei2025,Amiri2025}, Light Front Quark Model \cite{Zhao2018a}, QCD light cone sum rule \cite{Duan2025}, Model Independent Field Theory \cite{Sheng2020}, Relativistic Quark Model \cite{Ebert2006}, Large $N_c$ HQET \cite{Du2011,Du2013}, Constituent Quark Model (CQM) \cite{Pervin2006}, Bethe-Salpeter approach \cite{Ivanov1999}, Spectator Quark Model \cite{Singleton1991}, Bjorken sum rules \cite{Xu1993}, and HQET \cite{Sutherland1994}. Despite these efforts, notable discrepancies remain between the predicted form factors and decay widths, reflecting the different treatments of three-body dynamics, potential forms, and heavy quark symmetry-breaking corrections.

In this study, we extended our previous studies \cite{Thakkar2024,Thakkar2020,Patel2025a} on the exclusive semileptonic decay of the $\Lambda^0_b$ baryon. The main difference between this study and our previous studies is the method used to solve the six-dimensional Schr\"{o}dinger equation. In previous studies \cite{Thakkar2024,Thakkar2020,Patel2025a}, the Schr\"{o}dinger equation was solved variationally using a hyper-Coulomb trial wavefunction optimised through the virial theorem. In contrast, in the present study, it was solved numerically using the Runge-Kutta integration method to obtain the wavefunction\cite{Lucha1999}. This direct integration approach provides an exact numerical solution to the hyperradial equation for the chosen potential, thereby offering a higher accuracy and consistency. In this study, we investigated semileptonic decay $\Omega_b \rightarrow \Omega_c e \nu_e$ using the hypercentral constituent quark model (HCQM). We first compute the ground-state masses of $\Omega_b$ and $\Omega_c$ baryons in Section \ref{sec:2}, and then use the corresponding wavefunction to evaluate the Isgur-Wise function and the six independent form factors in Section \ref{sec:3}. The helicity amplitude formalism is employed in Section \ref{sec:4} to calculate the differential, total decay widths, and branching ratios. Our predictions are compared with the results of other theoretical approaches and discussed in detail in Section \ref{sec:5}. This study is summarised in Section \ref{sec:6}.

\section{Theoretical Framework}\label{sec:2}

We adopted the hypercentral constituent quark model to study the $\Omega_b^-$ baryon. We numerically solve the six dimensional Schr\"{o}dinger equation to obtain the masses of $\Omega_Q$ baryons using Wolfram Mathematica notebook \cite{Lucha1999}. In HCQM, Jacobi coordinates are essential for simplifying the three-body problem and providing a simplified representation of inter-quark dynamics. Jacobi coordinates provide the relevant degrees of freedom for the relative motion of the three constituent quarks and are given as follows
\begin{equation}\label{eq:1}
\boldsymbol{\rho} = \frac{1}{\sqrt{2}}(\mathbf{r}_1 - \mathbf{r}_2),
\end{equation}
\begin{equation}\label{eq:2}
\boldsymbol{\lambda} = \frac{{m_1 \mathbf{r}_1 + m_2 \mathbf{r}_2 -  (m_1 + m_2) \mathbf{r}_3}}{{\sqrt{m_1^2 + m_2^2 + (m_1 + m_2)^2}}}.
\end{equation}
The reduced masses are given as
\begin{equation}\label{eq:3}
m_{\rho}=\frac{2 m_{1} m_{2}}{m_{1}+ m_{2}},
\end{equation}
\begin{equation}\label{eq:4}
m_{\lambda}=\frac{2 m_{3} (m_{1}^2 + m_{2}^2+m_1m_2)}{(m_1+m_2)(m_{1}+ m_{2}+ m_{3})},
\end{equation}
where $m_1$, $m_2$ and $m_3$ are the constituent quark masses. The hyperspherical coordinates are given by the angles $\Omega_\rho=(\theta_\rho,\phi_\rho)$ and $\Omega_\lambda=(\theta_\lambda,\phi_\lambda)$. The hyper-radius $x$ and hyper-angle $\xi$ are defined as
\begin{equation}\label{eq:5}
x=\sqrt{\rho^2+\lambda^2};\xi=\arctan\left(\frac{\rho}{\lambda}\right).
\end{equation}
 Using hyperspherical coordinates, the kinetic energy operator $P_x^2/2m$ of the three-body system can be written as \cite{Thakkar2020}
\begin{equation}\label{eq:6}
\frac{P_x^2}{2m} = -\frac{1}{2m} \left(\frac{\partial^2}{\partial x^2} + \frac{5}{x} \frac{\partial}{\partial x} - \frac{L^2(\Omega_\rho,\Omega_\lambda,\xi)}{x^2}\right),
\end{equation}
where $m= \frac{2m_\rho m_\lambda}{m_\rho+m_\lambda}$ denotes reduced mass. $L^2(\Omega_\rho,\Omega_\lambda,\xi)$ is the quadratic Casimir operators of the six-dimensional rotational group $O(6)$ and its eigenfunctions are the hyperspherical harmonics $Y_{[\gamma]l_\rho l_\lambda}(\Omega_\rho,\Omega_\lambda,\xi)$ which satisfy the eigenvalue relation
\begin{equation}\label{eq:7}
L^2Y_{[\gamma]l_{\rho}l_{\lambda}}(\Omega_{\rho},\Omega_{\lambda},\xi)=\gamma(\gamma+4)Y_{[\gamma]l_{\rho}l_{\lambda}}(\Omega_{\rho},\Omega_{\lambda},\xi),
\end{equation}
where $l_\rho$ and $l_\lambda$ are the angular momenta associated with the $\rho$ and $\lambda$ variables respectively. The model Hamiltonian for baryons can be expressed as
\begin{equation}\label{eq:8}
H = \frac{P_\rho^2}{2m_\rho}+\frac{P_\lambda^2}{2m_\lambda}+V(\rho,\lambda) = \frac{P_x^2}{2m} + V(x).
\end{equation}
Here, the potential $V(x)$ is not purely a two-body interaction but also includes three-body effects. The six-dimensional hyperradial Schr\"{o}dinger equation can be written as
\begin{eqnarray}\label{eq:9}
\left[ -\frac{1}{2m}\frac{d^2}{dx^2} + \frac{\frac{15}{4}+\gamma(\gamma + 4)}{2mx^2} + V(x) \right] \phi_{\gamma}(x) = E\phi_{\gamma}(x),
\end{eqnarray}
where $\phi_{\nu\gamma} = x^{\frac{5}{2}}\psi_{\gamma}(x)$ is the hyperradial wave function and $\psi_{\gamma}(x)$ is the hypercentral wave function labelled by the grand angular quantum number $\gamma$ defined by the number of nodes $\nu$. The potential is assumed to depend only on the hyperradius and hence is a three-body potential because the hyperradius depends only on the coordinates of all three quarks. We consider the hypercentral potential $V(x)$ as the hyper-Coulomb (hC) plus linear potential, which is given as
\begin{equation}\label{eq:10}
V(x) = \frac{\tau}{x} + \beta x + V_0+ V_{spin},
\end{equation}\\
   where $\tau = -\frac{2}{3}\alpha_s$ is the hyper-Coulomb strength and the values of the potential parameters $\beta$ and $V_0$ are fixed to obtain the ground-state masses. $V_{spin}$ is the spin-dependent part that is perturbatively added and is given as \cite{Garcilazo2007,Patel2025,majethiya2008a}
\begin{equation}\label{eq:11}
V_{spin}(x) = -\frac{A}{4} \alpha_s  \frac{e^{-x/x_0}}{x {x_0}^2} \sum_{i<j} \boldsymbol{\lambda}_i \cdot \boldsymbol{\lambda}_j \frac{\boldsymbol{\sigma}_i \cdot \boldsymbol{\sigma}_j}{6 m_i m_j}.
\end{equation}
Here, parameter $A$ and regularisation parameter $x_0$ are considered as the hyperfine parameters of the model. Parameter $x_0$ is treated as a hyperfine parameter related to gluon dynamics, independent of the masses of the interacting quarks, as proposed in Ref. \cite{majethiya2008a}. As seen in Ref. \cite{majethiya2008a}, the values of $A$ vary depending on the quark content, which plays a crucial role in determining the mass splitting of single heavy baryons. There is no well-established procedure for evaluating $x_0$. $\boldsymbol{\lambda_{i,j}}$ are the SU(3) colour matrices, $\boldsymbol{\sigma_{i,j}}$ are the spin Pauli matrices, $m_{i,j}$ are the constituent masses of the two interacting quarks. Parameter $\alpha_s$ corresponds to the strong running coupling constant, which is given by
\begin{equation}\label{eq:12}
\alpha_s = \frac{\alpha_s(\mu_0)}{1+(\frac{33-2n_f}{12\pi})\alpha_s(\mu_0)\ln(\frac{m_1+m_2+m_3}{\mu_0})},
\end{equation}
     where $\alpha_s(\mu_0 = 1 GeV) \approx 0.6$ is considered in the present study. The masses of the ground-state $\Omega_Q$ baryons were calculated by summing the model quark masses and the binding energy.
\begin{equation}\label{eq:13}
M_B = m_1 + m_2 + m_3 + \langle H \rangle.
\end{equation}

\section{Form factors for $\Omega_b^-$ baryon decay}\label{sec:3}

For the semileptonic decay $\Omega_b^- \rightarrow \Omega_c^0\,e\,\bar{\nu_e}$, the effective Hamiltonian describing this transition can be written as

\begin{equation}\label{eq:14}
\mathcal{H}_{eff} = \frac{G_F}{\sqrt{2}} V_{cb}\, \bar{c}\gamma^{\mu}(1-\gamma_{5})b \, \bar{\ell}\gamma_{\mu}(1-\gamma_{5})\nu_{\ell},
\end{equation}
where $G_F$ is the Fermi coupling constant, and $V_{cb}$ denotes the relevant Cabibbo-Kobayashi-Maskawa (CKM) matrix element. The transition amplitude can be expressed as

\begin{eqnarray}\label{eq:15}
\mathcal{M} = \langle \Omega_c^{0} | \mathcal{H}_{eff} | \Omega_b^{-} \rangle \\ \nonumber
= \frac{G_F}{\sqrt{2}} V_{cb}\, \bar{\ell}\gamma^{\mu}(1-\gamma_{5})\nu_{\ell} \, \langle \Omega_c^{0} | \bar{c}\gamma_{\mu}(1-\gamma_{5})b | \Omega_b^{-} \rangle.
\end{eqnarray}

The hadronic matrix elements of the vector and axial-vector currents can be parameterised in terms of six form factors as follows \cite{Gutsche2015}
\begin{align}\label{eq:16}
M_{\mu}^{V} = \langle\Omega_{c}^{0}|\bar{c}\gamma_{\mu}b|\Omega_{b}^{-}\rangle = \bar{u}_{\Omega_{c}}(p^{\prime},\lambda^{\prime}) \\ \nonumber
\left[\gamma_{\mu}f_{1}(q^{2})-i\sigma_{\mu\nu}q^{\nu}f_{2}(q^{2})+q_{\mu}f_{3}(q^{2}) \right]u_{\Omega_{b}}(p,\lambda),\\ \nonumber
M_{\mu}^{A} = \langle\Omega_{c}^{0}|\bar{c}\gamma_{\mu}\gamma_{5}b|\Omega_{b}^{-}\rangle = \bar{u}_{\Omega_{c}}(p^{\prime},\lambda^{\prime}) \\ \nonumber
\left[ \gamma_{\mu}g_{1}(q^{2})-i\sigma_{\mu\nu}q^{\nu}g_{2}(q^{2})+q_{\mu}g_{3}(q^{2}) \right]\gamma_{5}u_{\Omega_{b}}(p,\lambda),
\end{align}

where $\sigma_{\mu\nu}=\frac{i}{2}[\gamma_\mu,\gamma_\nu]$. $\bar{u}_{\Omega_c}(p',\lambda')$ and $u_{\Omega_b}(p,\lambda)$ are the Dirac spinors of the $\Omega_c$ and $\Omega_b$ baryons, and $p^{(')}$ and $\lambda^{(')}$ are the corresponding momentum and helicity, respectively. Another parameterisation of these decay matrix elements can be expressed as follows
\begin{align}\label{eq:17}
M^{V}_{\mu} = \bar{u}_{\Omega_c}(p',\lambda')[ \gamma_{\mu}\,F_{1}(q^{2}) \\ \nonumber
+ F_{2}(q^{2})\,\frac{p_{\mu}}{M_{\Omega_b}} + F_{3}(q^{2})\,\frac{p'_{\mu}}{M_{\Omega_c}} ] u_{\Omega_b}(p,\lambda),\\ \nonumber
M^{A}_{\mu} = \bar{u}_{\Omega_c}(p',\lambda')\\ \nonumber
\left[ \gamma_{\mu}\,G_{1}(q^{2}) + G_{2}(q^{2})\,\frac{p_{\mu}}{M_{\Omega_b}}+ G_{3}(q^{2})\,\frac{p'_{\mu}}{M_{\Omega_c}} \right] \gamma_{5}\,u_{\Omega_b}(p,\lambda).
\end{align}

$f_i$ and $g_i$ with $i=1,2,3$ are the three form factors that describe the vector and axial vector transitions, respectively. The relationship between these two sets of form factors, Eq. (\ref{eq:16}) and (\ref{eq:17}) is expressed as in \cite{Falk1993}

\begin{eqnarray}\label{eq:18}
f_1 = F_1 + (m_{\Omega_b} + m_{\Omega_c}) \left( \frac{F_2}{2m_{\Omega_b}} + \frac{F_3}{2m_{\Omega_c}} \right), \nonumber \\
f_2 = -\frac{F_2}{2m_{\Omega_b}} - \frac{F_3}{2m_{\Omega_c}}, \nonumber \\
f_3 = \frac{F_2}{2m_{\Omega_b}} - \frac{F_3}{2m_{\Omega_c}}, \nonumber \\
g_1 = G_1 - (m_{\Omega_b} - m_{\Omega_c}) \left( \frac{G_2}{2m_{\Omega_b}} + \frac{G_3}{2m_{\Omega_c}} \right), \nonumber \\
g_2 = -\frac{G_2}{2m_{\Omega_b}} - \frac{G_3}{2m_{\Omega_c}}, \nonumber \\
g_3 = \frac{G_2}{2m_{\Omega_b}} - \frac{G_3}{2m_{\Omega_c}}.
\end{eqnarray}
In the framework of Heavy Quark Effective Theory (HQET), the structure of the form factors simplifies considerably in the infinite heavy-quark mass limit $m_Q\rightarrow\infty$ (Q = b,c). In this limit, heavy-quark spin symmetry implies that all form factors are reduced to a single universal Isgur-Wise function $\xi(\omega)$ \cite{Isgur1991}. \begin{eqnarray}\label{eq:21}
F_1(q^2) = G_1(q^2) = \xi(\omega),\nonumber \\
F_2 = F_3 = G_2 = G_3 = 0,
\end{eqnarray}

where $\omega=\upsilon\cdot\upsilon'$ is the velocity transfer between the initial $\upsilon$ and the final $\upsilon'$ heavy baryons. This is related to the squared four-momentum transfer between the heavy baryons, $q^2$ as $\omega=\frac{m^2_{\Omega_b}+m^2_{\Omega_c}-q^2}{2m_{\Omega_b}m_{\Omega_c}}$. The Isgur-Wise function $\xi(\omega)$ is expanded around zero recoil as \cite{Thakkar2020}

\begin{equation}\label{eq:22}
\xi(\omega)=1-\rho^2 (\omega-1)+c(\omega-1)^2+\ldots,
\end{equation}
where $\rho^2$ is the magnitude of the slope and $c$ is the curvature (convexity parameter) of the IWF ($\xi(\omega)$), which can be written in HCQM as in \cite{Thakkar2020,Patel2025a}
\begin{equation}\label{eq:23}
\rho^2=16 \pi^2 m^2\int_{0}^{\infty} |\psi_{\nu\gamma}(x)|^2 x^7 dx,
\end{equation}
\begin{equation}\label{eq:24}
c=\frac{8}{3} \pi^2 m^4\int_{0}^{\infty} |\psi_{\nu\gamma}(x)|^2 x^9 dx.
\end{equation}
Beyond the heavy-quark limit, subleading corrections of the order $1/m_Q$ arise in HQET \cite{Falk1993,Neubert1994}. These corrections originate from two sources. The first parameterises the $1/m_Q$ corrections to the heavy-quark current and is proportional to the product of the parameter $\bar{\Lambda}= m_{\Omega_b}-m{b}$, which is the difference between the baryon mass ($\Omega_b$) and heavy quark mass ($b$) in the infinitely heavy quark limit and the leading order Isgur-Wise function $\xi(\omega)$. The second originates from the kinetic energy term in the $1/m_Q$ correction to the HQET Lagrangian and introduces the additional function $A(\omega) = \frac{\bar{\Lambda}}{1+\omega}\xi(\omega)$ \cite{Georgi1990}. The baryon form factors in the HQET are expressed as \cite{Falk1993,Georgi1990,Neubert1994}
\begin{eqnarray}\label{eq:19}
F_1(\omega) = \xi(\omega) + \left( \frac{1}{2m_b} + \frac{1}{2m_c} \right) \left[ B_1(\omega) - B_2(\omega) \right], \nonumber \\
G_1(\omega) = \xi(\omega) + \left( \frac{1}{2m_b} + \frac{1}{2m_c} \right) B_1(\omega), \nonumber\\
F_2(\omega) = G_2(\omega) = \frac{1}{2m_c} B_2(\omega), \nonumber\\
F_3(\omega) = -G_3(\omega) = \frac{1}{2m_b} B_2(\omega),
\end{eqnarray}
where the functions $B_1$ and $B_2$ are expressed as \cite{Falk1993}
\begin{eqnarray}\label{eq:20}
B_1(\omega) = \bar{\Lambda} \frac{w - 1}{w + 1} \, \xi(\omega) + A(\omega), \nonumber\\
B_2(\omega) = -\frac{2\bar{\Lambda}}{w + 1} \, \xi(\omega).
\end{eqnarray}

\section{Helicity Amplitudes and Heavy-to-Heavy \\Semileptonic decay of $\Omega_b$ Baryon }\label{sec:4}

The helicity amplitudes are expressed in terms of the baryon form factors as \cite{Bialas1993}

\begin{eqnarray}\label{eq:26}
H^{V,A}_{+\frac{1}{2},0} = \frac{1}{\sqrt{q^2}} \sqrt{2M_{\Omega_b}M_{\Omega_c}(\omega\mp1)} \\ \nonumber
\times[(M_{\Omega_b}\pm M_{\Omega_c}) \mathcal{F}^{V,A}_1(\omega) \pm M_{\Omega_c}(\omega\pm1) \mathcal{F}^{V,A}_2(\omega)\\ \nonumber
\pm M_{\Omega_b}(\omega\pm1) \mathcal{F}^{V,A}_3(\omega)],
\end{eqnarray}
\begin{eqnarray}\label{eq:27}
H^{V,A}_{+\frac{1}{2},1} = -2 \sqrt{M_{\Omega_b}M_{\Omega_c}(\omega\mp1)} \mathcal{F}^{V,A}_1(\omega),
\end{eqnarray}
\begin{eqnarray}\label{eq:28}
H^{V,A}_{+\frac{1}{2},t} = \frac{1}{\sqrt{q^2}} \sqrt{2M_{\Omega_b}M_{\Omega_c}(\omega\pm1)} \\ \nonumber
\times[(M_{\Omega_b}\mp M_{\Omega_c}) \mathcal{F}^{V,A}_1(\omega)\pm (M_{\Omega_b}-M_{\Omega_c} \omega) \mathcal{F}^{V,A}_2(\omega)\\ \nonumber
\pm (M_{\Omega_b} w-M_{\Omega_c}) \mathcal{F}^{V,A}_3(\omega)],
\end{eqnarray}

where the superscripts correspond to vector (V) and axial-vector (A), and $\mathcal{F}^V_i(\omega)\equiv F_i(i = 1,2,3)$, $\mathcal{F}^A_i(\omega)\equiv G_i(i = 1,2,3)$. $H^{V,A}_{\lambda',\lambda_W}$ are the helicity amplitudes for weak transitions induced by the vector (V) and axial-vector (A) currents, where $\lambda'$ and $\lambda_W$ are the helicities of the final baryon and virtual $W$ boson, respectively. $\lambda_W = 0,\pm1$ for the vector case and $\lambda_W = 0$ for the scalar case. To distinguish between the two $\lambda_W = 0$ states, the scalar is typically written as $\lambda_W = t$ which is temporal. This scalar does not contribute to semileptonic decay in the massless lepton limit. The helicity-flipped (negative) amplitudes can be obtained using the relations $H^{V}_{-\lambda',-\lambda_W} = H^{V}_{\lambda',\lambda_W}$ and $H^{A}_{-\lambda',-\lambda_W} = -H^{A}_{\lambda',\lambda_W}$ \cite{Faustov2016}. The total helicity amplitude for the V-A current is given by

\begin{equation}\label{eq:29}
H_{\lambda',\lambda_W} = H^V_{\lambda',\lambda_W} - H^A_{\lambda',\lambda_W}.
\end{equation}

The helicity structure functions with definite parity can be expressed in terms of bilinear combinations of the helicity amplitudes. The relevant parity-conserving helicity structures are expressed as \cite{Gutsche2015}

\begin{eqnarray}\label{eq:30}
\mathcal{H}_T(q^2) &= |H_{+\frac{1}{2},1}|^2 + |H_{-\frac{1}{2},-1}|^2, \\ \nonumber
\mathcal{H}_L(q^2) &= |H_{+\frac{1}{2},0}|^2 + |H_{-\frac{1}{2},0}|^2, \\ \nonumber
\mathcal{H}_S(q^2) &= |H_{+\frac{1}{2},t}|^2 + |H_{-\frac{1}{2},t}|^2,
\end{eqnarray}
where $\mathcal{H}_T(q^2)$ is the transverse unpolarised helicity, $\mathcal{H}_L(q^2)$ is the longitudinal unpolarised helicity and $\mathcal{H}_S(q^2)$ is the scalar unpolarised helicity. The differential decay rate is expressed as follows \cite{Gutsche2015,Faustov2016,Migura2006}

\begin{equation}\label{eq:31}
\frac{d\Gamma}{dq^2} = \frac{G_F^2}{8\pi^3}|V_{cb}|^2 \frac{\lambda^{\frac{1}{2}}(q^2-m_l^2)^2}{48M_{\Omega_b}^3 q^2} \mathcal{H}_{Total}(q^2),
\end{equation}
where $G_F = 1.16\times10^{-5} GeV^{-2}$ is the Fermi coupling constant, $|V_{cb}|= 0.041$ is the CKM matrix element, $\lambda = [(M_{\Omega_b}^2 - M_{\Omega_c}^2) - q^2]^2 $. $m_l$ is the lepton mass ($l=e,\mu,\tau$). The total helicity is defined as follows

\begin{eqnarray}\label{eq:32}
\mathcal{H}_{Total}(q^2) = \mathcal{H}_T(q^2) + \mathcal{H}_L(q^2)\\ \nonumber
+\frac{m_l^2}{2q^2} (\mathcal{H}_T(q^2) + \mathcal{H}_L(q^2) + 3\mathcal{H}_S(q^2)),
\end{eqnarray}

where the first two terms are non-spin-flip contributions and the last three terms proportional to $m_l$ are lepton-helicity flip contributions. By integrating the differential decay rate in Eq. (\ref{eq:31}) ($q^2 \in [m^2_e, (M_{\Omega_b} - M_{\Omega_c})^2]$), we obtain the total decay rate.

\section{Results and discussions}\label{sec:5}

The baryon masses were obtained by numerically solving the six-dimensional Schr\"{o}dinger equation within the hypercentral constituent quark model (HCQM), employing a potential composed of a hyper-Coulomb like term and a linear term. A spin-dependent interaction is included phenomenologically to reproduce the observed mass splittings among baryon states. The hypercentral potential given in Eq. (\ref{eq:10}) provides an effective description of the collective three-quark dynamics in terms of the hyperradius $x$. Since the potential depends only on $x$, it offers a collective parametrisation of the three-body system. Although it does not reproduce the exact pairwise QCD interactions, it serves as an effective description of baryon dynamics within a three-body framework. This approach has been widely used in baryon spectroscopy and has demonstrated reasonable success in describing ground-state and excited-state baryon masses, magnetic moments, radiative decays, and semileptonic transitions.\\
In this study, we employed the quark mass parameters listed in Table \ref{tab:table1} to calculate the ground-state masses of the $\Omega_b$ and $\Omega_c$ baryons in the HCQM. The predicted masses of the $\Omega_b$ and $\Omega_c$ baryons are $\Omega_b$ = 6.045 GeV and $\Omega_c$ = 2.689 GeV, respectively, which are in good agreement with the experimental results reported by the Particle Data Group \cite{pdg2024}. This agreement demonstrates the reliability of the HCQM framework for describing heavy baryons and provides a consistent foundation for evaluating the semileptonic transition of $\Omega_b$ baryons. \\
\begin{table}
\centering
  \caption{\label{tab:table1}Quark mass parameters and constants used in the calculations.}
   \begin{tabular}{cc}
    \noalign{\smallskip}\hline
    Parameter & Value\\
    \noalign{\smallskip}\hline
    ${m_{s}}$ & 0.50 GeV\\
    ${m_{c}}$ & 1.55 GeV\\
    ${m_{b}}$ & 4.95 GeV\\
    $\beta$ & 0.2 $GeV^2$\\
    ${V_0}$ for bottom baryons & -0.994 GeV\\
    ${V_0}$ for charmed baryons &-0.930GeV\\
    A for bottom baryons & 1 \\
    A for charmed baryons & 25 \\
    $x_0$ & 1.00 $GeV^{-1}$\\
    $\alpha_s(\mu_0 = 1 GeV)$ & 0.6\\
    \noalign{\smallskip}\hline
    \end{tabular}
\end{table}
Such decays are described by hadronic form factors. Using HQET in the subleading order, these form factors are related to the universal Isgur-Wise function $\xi(\omega)$, which is generally an overlap integral involving the final and initial wavefunctions. This is because, near the zero recoil point ($\omega=1$), the four velocities of the baryons before and after the transitions are identical. The slope $\rho^2$ and the convexity parameter $c$ obtained using the wavefunction (see Eq. (\ref{eq:23}) and (\ref{eq:24})) are 3.43 and 2.04, respectively. These values are slightly larger than those reported in Ref. \cite{Hassanabadi2014} ($\rho^2$ = 2.56 and $c$ = 1.43), which may arise from the choice of potential parameters and wavefunction normalization in HCQM. The slope reported in Ref. \cite{Ivanov1997} was 2.79. The Isgur-Wise function was constructed using the slope and convexity parameters through Eq. (\ref{eq:20}) and subsequently employed to compute the six form factors.

\begin{figure*}
\includegraphics[scale=0.33]{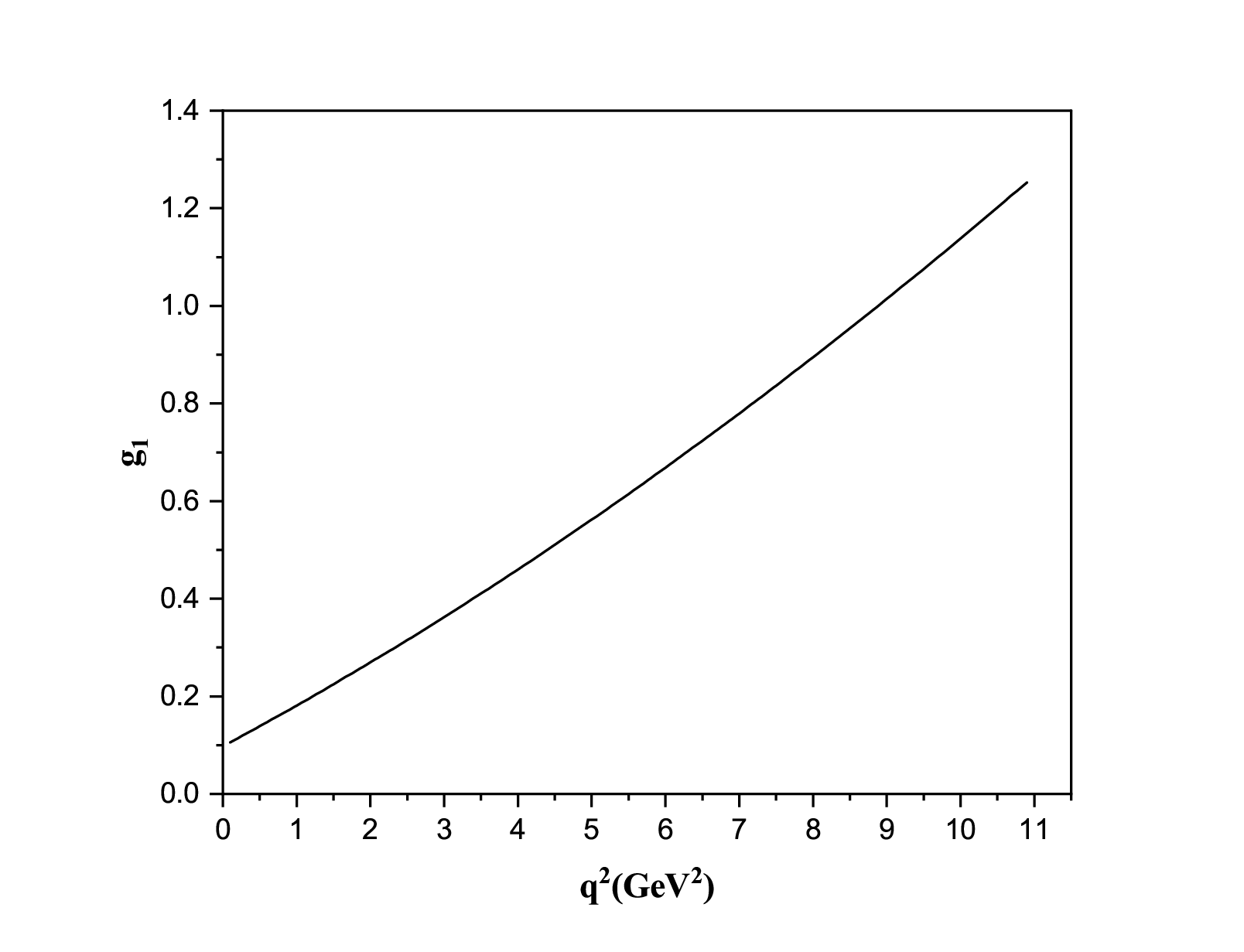}
\includegraphics[scale=0.33]{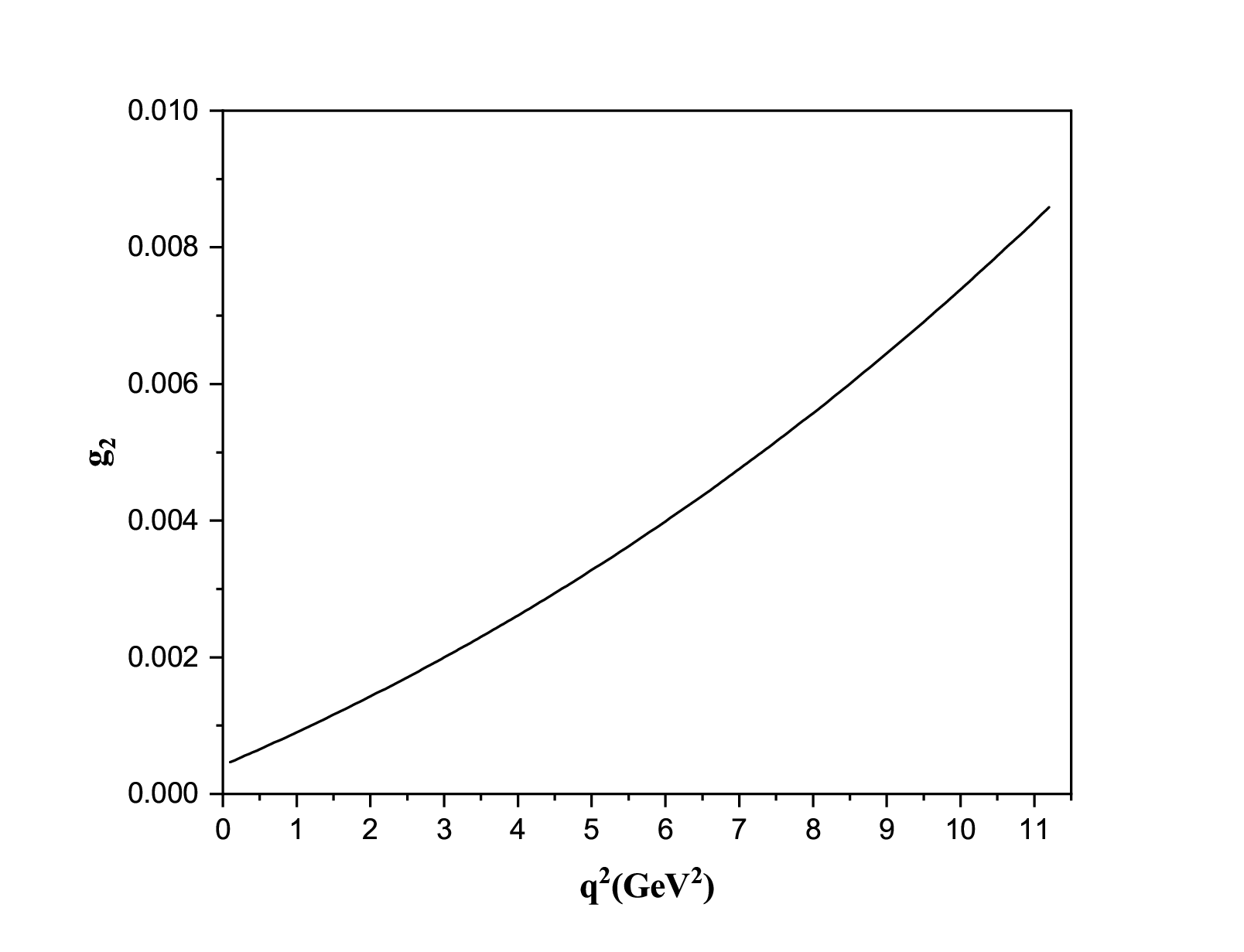}
\includegraphics[scale=0.33]{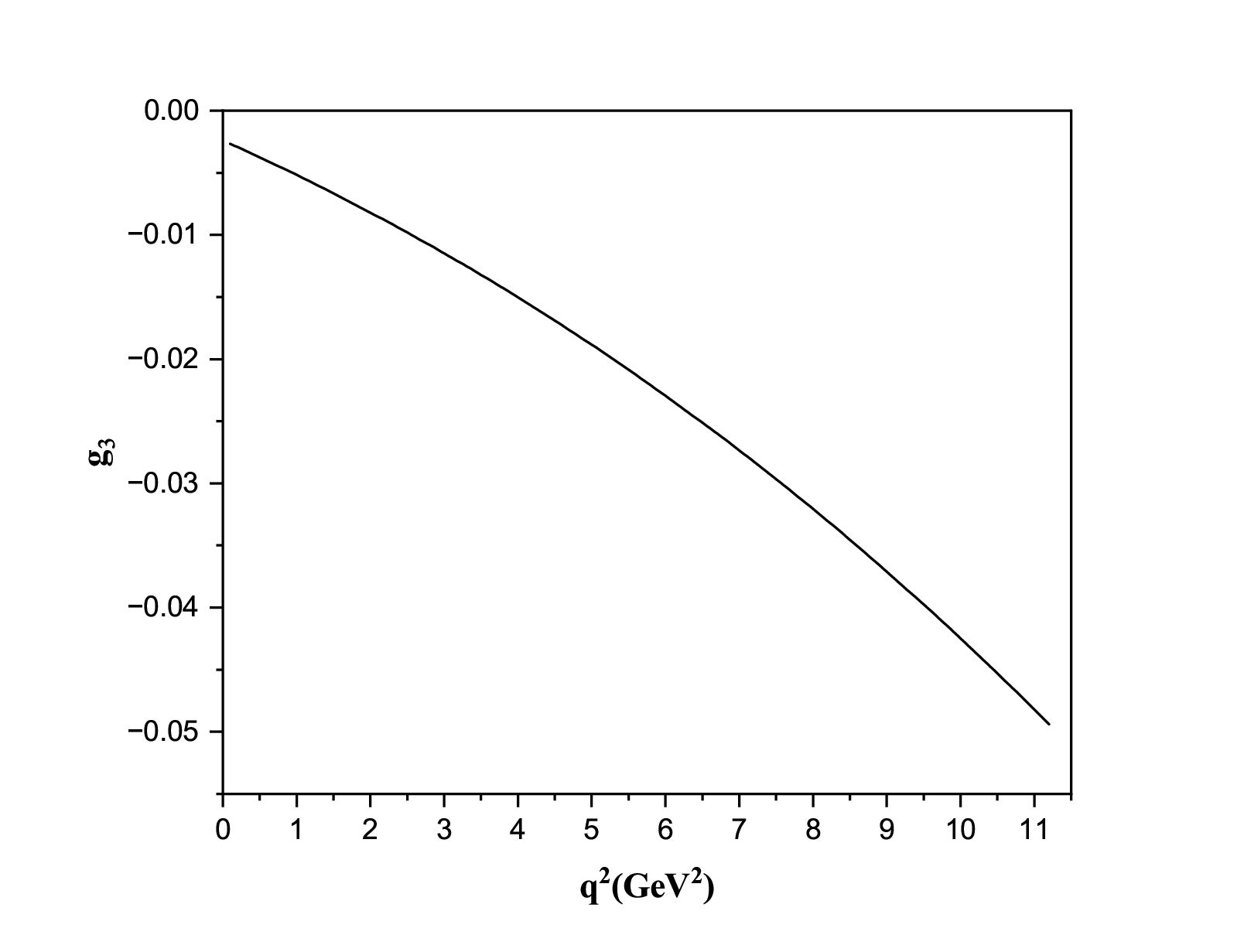}
\caption{\label{fig:1} The form factors $f_1$, $f_2$, $f_3$, $g_1$, $g_2$ and $g_3$ as a function of $q^2$. }
\end{figure*}

\begin{figure}
\includegraphics[scale=0.4]{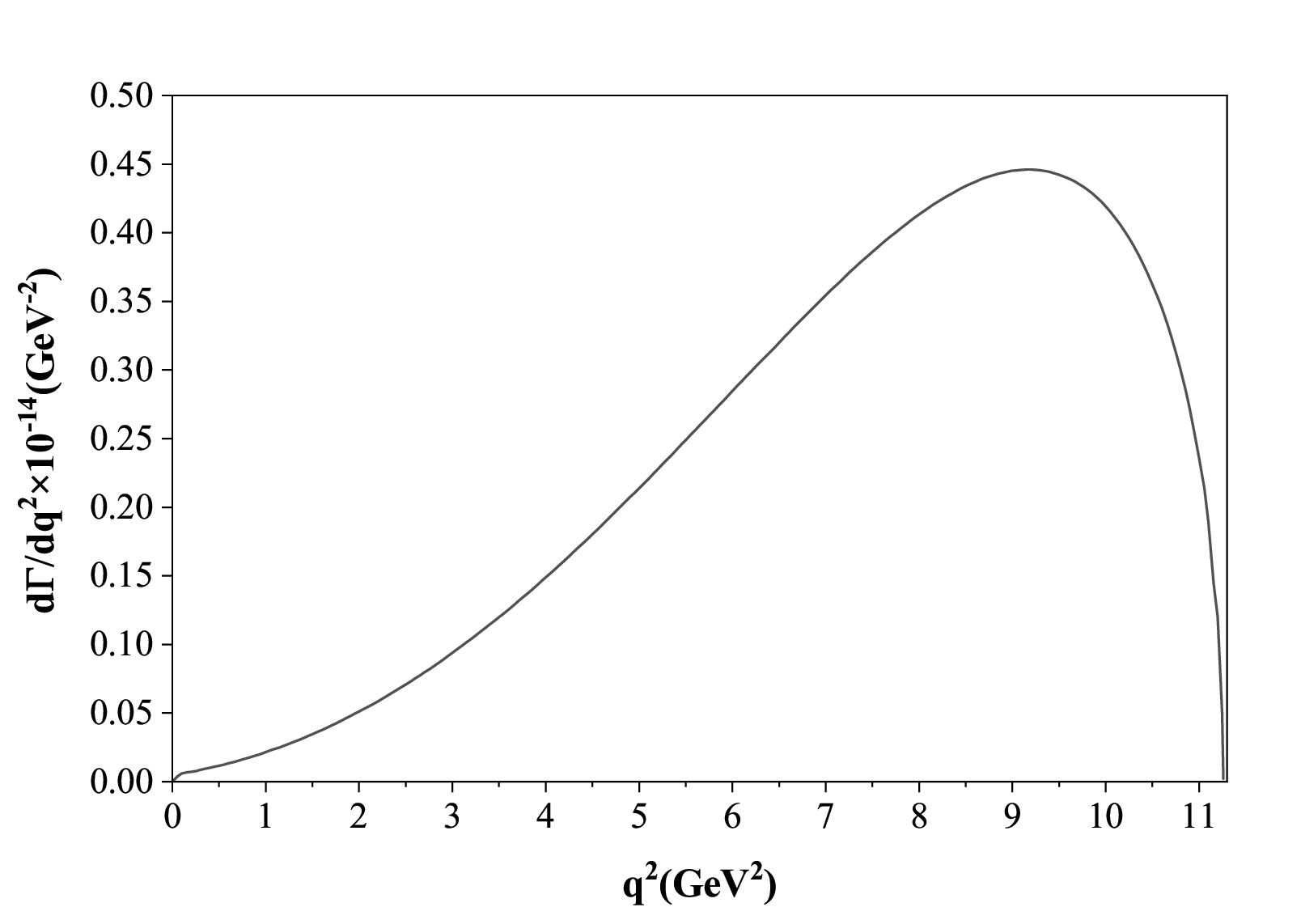}
\caption{\label{fig:2} The semileptonic decay of the $\Omega_b^- \rightarrow \Omega_c^0\,e\,\bar{\nu_e}$ in $q^2$ kinetic region. }
\end{figure}

Among the six form factors, $F_1$ and $G_1$ dominate the kinematic region, indicating their direct association with the vector and axial currents, respectively. It is observed that the form factors $f_1$ and $g_1$ exhibit nearly identical numerical values. This behaviour can be understood within the framework of HQET, where both form factors are dominated by the leading-order Isgur-Wise function $\xi(\omega)$, which governs the heavy-to-heavy transition in the infinite heavy-quark mass limit. Although corrections up to order $1/m_Q$ are included in the present analysis, these contributions are subleading and suppressed by the inverse heavy quark masses. Consequently, their numerical impact remains small, leading to nearly equal values of $f_1$ and $g_1$.
The subleading form factors $F_2$, $F_3$, $G_2$, and $G_3$ were suppressed by factors of $1/m_Q$, as expected in HQET (see Table \ref{tab:table2}). The $q^2$ behaviours of the six form factors in the allowed regions are illustrated in Fig. \ref{fig:1}. In the strict heavy quark limit, all form factors are reduced to a universal Isgur-Wise function. The small but nonvanishing values of $F_2$, $F_3$, $G_2$ and $G_3$ arise from $1/m_b$ and $1/m_c$ corrections, which are parameterised in our formalism through $\bar{\Lambda}/m_Q$. Zhao et al. \cite{Zhao2018a}, working in the light-front quark model, reported only $f_{1,2}$, $g_{1,2}$. As discussed in their work, additional form factors $f_3$ and $g_3$ cannot be extracted in their $q^+ = 0$ framework, whereas our HCQM approach provides access to all six form factors. The reported values of the form factors at $q^2 = 0$ are $f_1 = 0.566$, $f_2 = 0.531$, $g_1 = -0.170 $ and $g_2 = -0.031$ \cite{Zhao2018a}. Significant quantitative differences were observed between the two approaches. In particular, the leading form factor $f_1$ obtained in the present HCQM framework is significantly smaller than that reported in the light-front quark model, and similar deviations are seen for the subleading form factors. As mentioned previously, Ref. \cite{Zhao2018a} employs a light-front quark model with a diquark treatment of the two spectator quarks whereas the present work is based on a hypercentral framework. The model dependence (HCQM versus the light-front quark model), different wavefunction shapes and the treatment of quark dynamics can significantly affect the magnitude and momentum dependence of the form factors. Despite the numerical differences, both calculations agree on the hierarchy $f_{1}, g_{1} \gg f_{2,3}, g_{2,3}$ consistent with the HQET expectations. More theoretical work is required on these form factors to further enrich our knowledge of the weak decay.
\begin{table}[h]
\centering
   \caption{\label{tab:table2} Form factors for $\Omega_b^- \rightarrow \Omega_c^0\,e\,\bar{\nu_e}$ decay}
    \begin{tabular}{cccc}
    \noalign{\smallskip}\hline
    form factor & at $q^2=0$ & at $q^2_{max}$ \\
    \noalign{\smallskip}\hline
    $f_1$ 	&	0.073	&	1.261	\\
    $f_2$ 	&	0.0024	&	0.05	\\
    $f_3$ 	&	-0.0004	&	-0.008	\\
    $g_1$ 	&	0.073	&	1.261	\\
    $g_2$ 	&	0.0004	&	0.008	\\
    $g_3$ 	&	-0.0024	&	-0.05	\\
    \noalign{\smallskip}\hline
    \end{tabular}
\end{table}
Using the computed form factors within the helicity formalism, we evaluate the differential and total semileptonic decay widths of the $\Omega_b^- \rightarrow \Omega_c^0\,e\,\bar{\nu_e}$ transition. The obtained heavy-to-heavy semileptonic decay rate and branching ratio for the $\Omega_b^- \rightarrow \Omega_c^0\,e\,\bar{\nu_e}$ transition were compared with the available theoretical predictions in Table \ref{tab:table3}. The differential decay width increases with increasing $q^2$, peaking at approximately $q^2 \approx 9.5$ GeV$^2$ and vanishing at $q^2_{max} = 11.2$ GeV$^2$ (Fig. \ref{fig:2}). The predicted decay width is $4.01 \times 10^{10} sec^{-1}$, corresponding to a branching ratio of 6.57\%. The value for the semileptonic decay rate reported by various theoretical predictions lies between 0.71 $\times$ $10^{10} sec^{-1}$ and 5.4 $\times$ $10^{10} sec^{-1}$. Our result is in agreement with models such as the Spectator Quark Model \cite{Singleton1991} and the Large $N_c$ HQET\cite{Du2013}. The behaviour of the differential decay rate in the $q^2$ region is shown in Fig. \ref{fig:2}. The branching ratio is calculated using the experimental lifetime of the $\Omega_b$ baryon listed in PDG \cite{pdg2024} as $\tau= 1.64 \times 10^{-12} sec$ and the decay widths listed in Table \ref{tab:table3}. As previously mentioned, no experimental data were available for this decay channel. However, the production of the $\Omega_b^-$ baryon spectra has been observed, and in the near future, such decay modes will be observed experimentally.
\begin{figure*}
\includegraphics[scale=0.33]{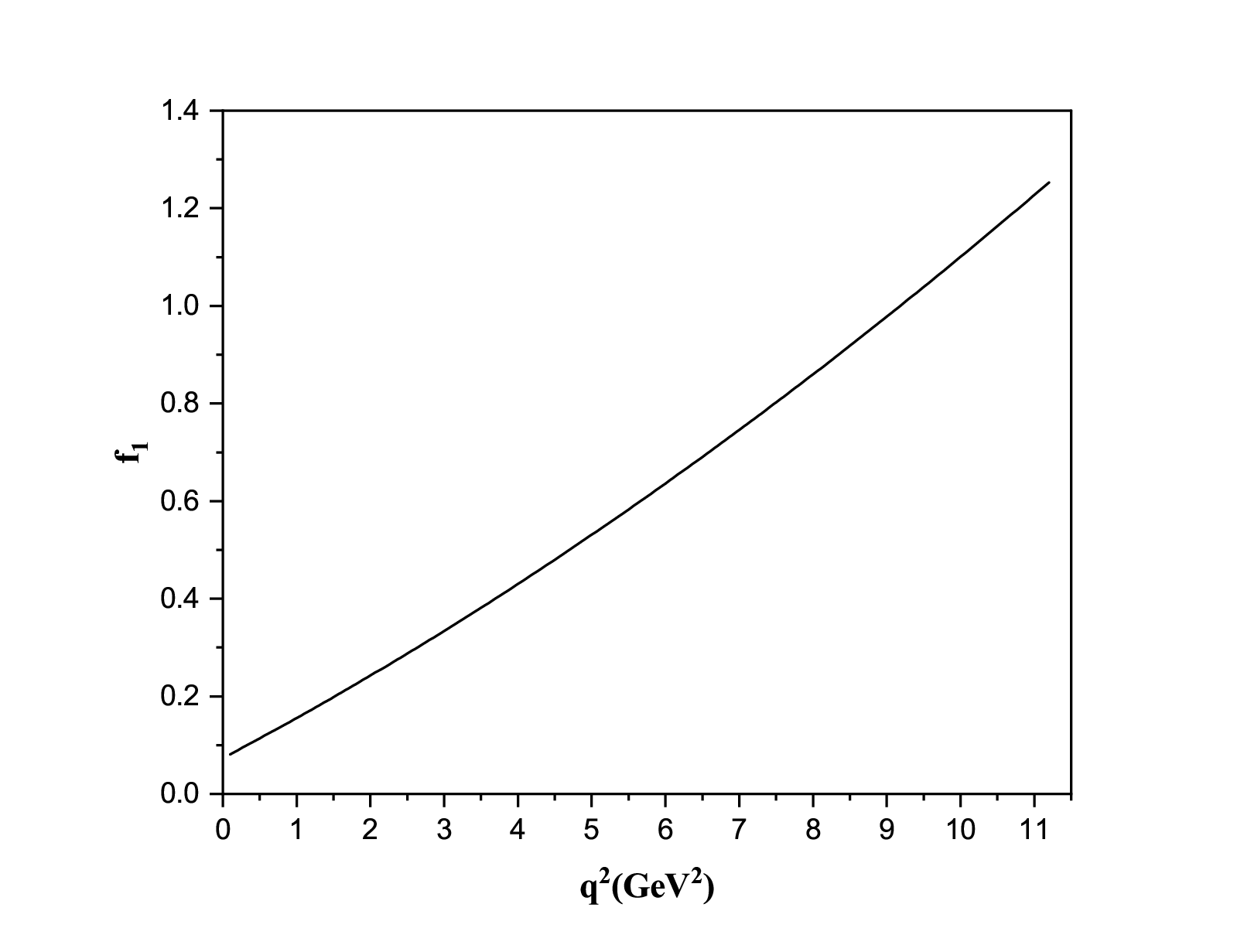}
\includegraphics[scale=0.33]{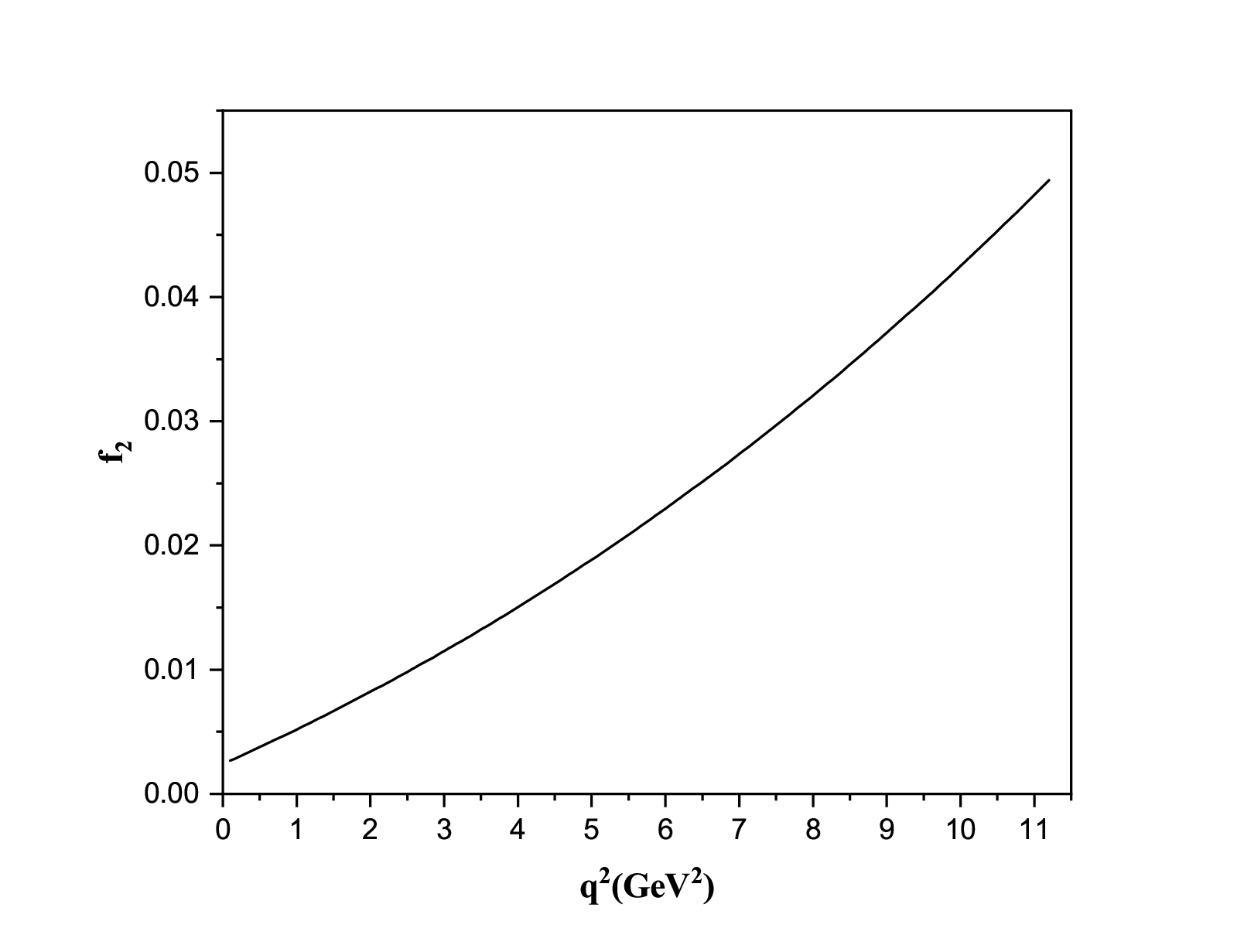}
\includegraphics[scale=0.33]{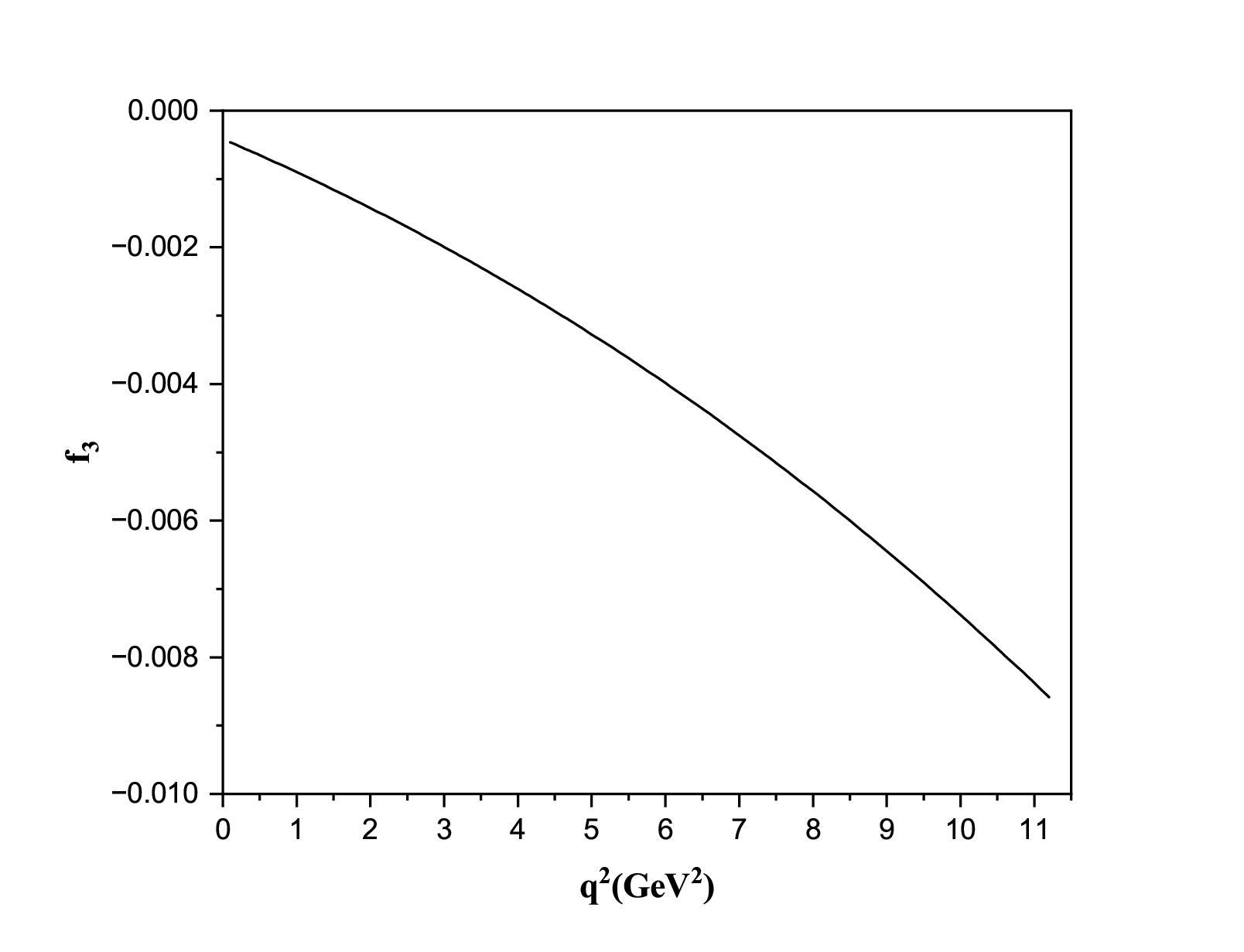}
\end{figure*}

\begin{table*}[h]
\centering
   \caption{{\label{tab:table3} Semileptonic decay width (in $10^{10} s^{-1}$) and Branching ratio (in \%) of $\Omega_b^- \rightarrow \Omega_c^0\,e\,\bar{\nu_e}$ transition}}
    \begin{tabular}{llll}
    \noalign{\smallskip}\hline
    Ref.	&	Decay Width	&	Br (\%)	 \\
    \noalign{\smallskip}\hline
    Our	&	4.01	&	6.57	\\
    Spectator quark model \cite{Singleton1991}	&	5.40	&	8.86	\\
    Large $N_c$ HQET \cite{Du2013}	&	5.13	&	8.41	\\
    Covariant quasi potential approach \cite{Rusetsky1997}	&	2.62	&	4.29	 \\
    Large $N_c$ HQET \cite{Du2011}	&	2.56	&	4.20	\\
    Bethe-Salpeter approach (1) \cite{Ivanov1999}	&	2.05	&	3.36	\\
    Bethe-Salpeter approach (2) \cite{Ivanov1999}	&	1.81	&	2.97	\\
    $1/m_Q$ correction HQET \cite{Xu1993a}	&	2.01	&	3.29	\\
    Non relativistic quark model \cite{Cheng1996}	&	2.30	&	3.77	\\
    quark-diquark model \cite{Farhadi2023}	&	$1.82\pm0.12$	& 	 $2.98\pm0.20$	 \\
    Relativistic three quark model \cite{Ivanov1997}	&	1.87	&	3.07	 \\
    QCD Sum Rule \cite{Neishabouri2024}	&	$1.67^{+1.00}_{-0.79}$	&	 $2.74^{+0.16}_{-0.13}$	\\
    HCQM \cite{Hassanabadi2014}	&	1.55	&	2.54	\\
    Relativistic in quasi potential approach \cite{Ebert2006}	&	1.29	&	 2.12	\\
    Independent model \cite{Sheng2020}	&	1.295	&	2.12	\\
    Relativistic three quark model \cite{Ivanov2000}	&	1.86	&	3.05	 \\
    light-front approach \cite{Zhao2018a}	&	1.73	&	2.84	\\
    HQET \cite{Korner1994}   &  1.52  &  2.49  \\
    QCD Sum Rule \cite{Lu2026} & $0.978^{+0.334}_{-0.234}$ & $2.44^{+0.83}_{-0.58}$ \\
    Constituent Quark Model \cite{Pervin2006}	&	0.71	&	1.16 \\
    \noalign{\smallskip}\hline
    \end{tabular}
\end{table*}
\section{Conclusion}\label{sec:6}

In this study, we investigated the semileptonic decay process $\Omega_b^- \rightarrow \Omega_c^0\,e\,\bar{\nu_e}$ within the framework of the Hypercentral Constituent Quark Model (HCQM). The six dimensional Schr\"{o}dinger equation was solved numerically using a hyper-Coulomb plus linear confining potential, perturbatively including the spin-dependent interactions. The obtained ground-state masses of the initial and final baryons showed good agreement with the experimental values, validating the reliability of HCQM in describing heavy baryon systems. The Isgur-Wise function and its slope and curvature parameters were extracted from the hyperradial wavefunction, which were subsequently used to compute the six independent form factors $F_i$ and $G_i$ at the subleading order of the Heavy Quark Effective Theory (HQET). Helicity amplitudes were obtained using these form factors to further evaluate the heavy-to-heavy semileptonic decay rate of $\Omega_b^-$ baryon. The branching ratios were computed and compared using various approaches. \\




\begin{thebibliography}{99}

\bibitem{pdg2024}S. Navas et al. (Particle Data Group), Phys. Rev. D 110, 030001 (2024).
\bibitem{Abazov2008} V. M. Abazov et al. (D0 Collaboration), Phys. Rev. Lett. 101, 232002 (2008).
\bibitem{Aaltonen2009} T. Aaltonen et al. (CDF Collaboration), Phys. Rev. D 80, 072003 (2009).

\bibitem{Neishabouri2024}Z. Neishabouri, K. Azizi, H. R. Moshfegh, Phys. Rev. D 110(1), 014010 (2024).
\bibitem{Khajouei2025}L. Khajouei, K. Azizi, arXiv:2510.01429 [hep-ph] (2025).
\bibitem{Amiri2025} A. Amiri, P. Eslami, K. Azizi, R. Jafari, arXiv:2509.01195v1 [hep-ph] (2025).
\bibitem{Zhao2018a}	Z. X. Zhao, Chin. Phys. C 42, 9, 093101 (2018).
\bibitem{Duan2025} Hui-Hui Duan, Yong-Lu Liu, Qin Chang, Ming-Qiu Huang, Phys. Rev. D 112, 054030 (2025).
\bibitem{Sheng2020}	J. H. Sheng, J. Zhu, X. N. Li, Q. Y. Hu, R. M. Wang, Phys. Rev. D 102(5), 055023 (2020).
\bibitem{Ebert2006}	D. Ebert, R. N. Faustov, V. O. Galkin, Phys. Rev. D 73, 094002 (2006).
\bibitem{Du2011}M. K. Du and C. Liu, Phys. Rev. D 84, 056007 (2011).
\bibitem{Du2013} M. K. Du, C. Liu, Phys. Rev. D 87 (9) 094015 (2013).
\bibitem{Pervin2006}M. Pervin, W. Roberts, S. Capstick, Phys. Rev. C 74, 025205 (2006).
\bibitem{Ivanov1999} M. A. Ivanov, J. G. K\"{o}rner, V. E. Lyubovitskij, A. G.Rusetsky, Phys. Rev. D 59, 074016 (1999).
\bibitem{Singleton1991}	R. Singleton, Phys. Rev. D 43, 2939 (1991).
\bibitem{Xu1993} Q. P. Xu, Phys. Rev. D 48, 5429-5432 (1993).
\bibitem{Sutherland1994}M. Sutherland, Z. Phys. C 63, 111 (1994).

\bibitem{Thakkar2020}K. Thakkar, Eur. Phys. J. C 80 (10), 926 (2020).
\bibitem{Thakkar2024}K. Thakkar, Indian J. Phys. 98 1, 365-369 (2024).
\bibitem{Patel2025a}K. Patel and K. Thakkar, Eur. Phys. J. Plus 140, 452 (2025).
\bibitem{Lucha1999}W. Lucha, F.F. Schoberl, Int. J. Mod. Phys. C 10, 607 (1999)

\bibitem{Garcilazo2007} H. Garcilazo, J. Vijande and A. Valcarce, J. Phys. G 34, 961 (2007).
\bibitem{Patel2025}K. Patel, K. Thakkar, Int. J. Theor. Phys. 64, (5) 129 (2025).
\bibitem{majethiya2008a}A. Majethiya, B. Patel and P. C. Vinodkumar, Eur. Phys. J. A 38, 307 (2008).

\bibitem{Gutsche2015}T. Gutsche, M. A. Ivanov, J. G. K\"{o}rner, V. E. Lyubovitskij, P. Santorelli and N. Habyl, Phys. Rev. D 91, 074001 (2015).
\bibitem{Falk1993} A. Falk, M. Neubert, Phys. Rev. D 47, 2982 (1993).
\bibitem{Georgi1990} H. Georgi, B. Grinstein, and M. B. Wise, Phys. Lett. B 252, 456 (1990).
\bibitem{Neubert1994}M. Neubert, Phys. Rep. 245, 259 (1994).
\bibitem{Isgur1991}N. Isgur and M. B. Wise, Nucl. Phys. B 348, 276 (1991).

\bibitem{Hassanabadi2014}	H. Hassanabadi, S. Rahmani, S. Zarrinkamar, Phys. Rev. D 90, 074024 (2014).
\bibitem{Bialas1993}P. Bialas, J. G. K\"{o}rner, M. K\"{a}rmer, K. Zalewski, Z. Phys. C 57, 115 (1993).
\bibitem{Faustov2016} R. N. Faustov and V. O. Galkin, Phys. Rev. D 94, 7, 073008 (2016).
\bibitem{Migura2006} S. Migura, D. Merten, B. Metsch and H. R. Petry, Eur. Phys. J. A 28, 55 (2006).
\bibitem{Rusetsky1997}	A. G. Rusetsky, M. A. Ivanov, J. G. K\"{o}rner, V. E. Lyubovitskij, arXiv:hep-ph/9710524 [hep-ph].

\bibitem{Farhadi2023} M. Farhadi et al., Eur. Phys. J. A 59, 171 (2023).
\bibitem{Ivanov1997} M. A. Ivanov, V. E. Lyubovitskij, J. G. K\"{o}rner, P. Kroll, Phys. Rev. D 56, 348 (1997).

\bibitem{Xu1993a}Q. P. Xu, A. N. Kamal, Phys. Rev. D 47, 2849-2857 (1993).
\bibitem{Cheng1996}	H. Y. Cheng and B. Tseng, Phys. Rev. D 53, 1457 (1996) [erratum: Phys. Rev. D 55, 1697 (1997)].
\bibitem{Ivanov2000}M. A. Ivanov, J. G. K\"{o}rner, V. E. Lyubovitskij, M. A. Pisarev and A. G. Rusetsky, Phys. Rev. D 61, 114010 (2000).

\bibitem{Korner1994}J. G. K\"{o}rner, D. Pirjol, M. K\"{a}rmer, Prog. Part. Nucl. Phys. 33, 787 (1994).
\bibitem{Lu2026} J. Lu, G.-L. Yu, D.-Y. Chen, Z.-G. Wang, B. Wu, arXiv:2602.04311 [hep-ph]

\end{thebibliography}
\end{document}